# Forced Synchronization of Spaser by an External Optical Wave

E.S. Andrianov,[1,2] A.A. Pukhov,[1,2] A.V. Dorofeenko,[1,2] A.P. Vinogradov,[1,2] and A.A. Lisyansky[3]

[1]Moscow Institute of Physics and Technology, 9 Institutskiy per., Dolgoprudniy 141700, Moscow Reg., Russia

[2]Institute for Theoretical and Applied Electromagnetics, 13 Izhorskaya, Moscow 125412, Russia

[3]Department of Physics, Queens College of the City University of New York, Flushing, NY 11367

We demonstrate that when the frequency of the external field differs from the lasing frequency of an autonomous spaser, the spaser exhibits stochastic oscillations at low field intensity. The plasmon oscillations lock to the frequency of the external field only when the field amplitude exceeds a threshold value. We find a region of values of the external field amplitude and the frequency detuning (the Arnold tongue) for which the spaser synchronizes with the external wave.

Joule (ohmic) losses in metallic nano-inclusions embedded in a host metamaterial medium are the main obstacle to realizing a practical "perfect" lens with subwavelength resolution as well as to other applications of metamaterials [1]. Losses can be compensated for, however, in amplifying media. The authors of Refs. [2, 3] proposed making a stackable lens with alternating metamaterial and gain layers This can be accomplished by directly introducing gain inclusions such as molecules or quantum dots (QDs) into the matrix [4-6]. The latter scheme has recently been realized experimentally [7-10] demonstrating that it is possible to compensate for losses.

The combination of a gain medium and metallic nanoparticles (NPs) results in the emergence of a spaser (Surface Plasmon Amplification by Stimulated Emission of Radiation) first suggested by Bergman and Stockman [11] and realized experimentally in Ref. [12]. Schematically, the spaser is a system of inversely excited two-level QDs surrounding metal nanoparticles [11, 13]. Its principles of operation are analogous to those for a laser with the role of photons played by surface plasmons (SPs) localized at a NP that serves as the resonator [11, 14, 15]. In other words, in a spaser, near-fields of the NP are generated and amplified. The amplification of the SPs occurs due to non-radiative energy transfer from QDs. This process takes place due to the dipole-dipole (or any other near-field [16]) interaction between the QD and the plasmon NP. The efficiency of such a mechanism depends on the probability of the non-radiative excitation of the surface plasmon, which is $(kr)^{-3}$ greater than the radiation of the photon [17], where $r$ is the distance between the centers of the QD and the NP and $k$ is the photon wavenumber. Stimulated radiation from the QD into the plasmon mode results in spasing. The excitation of the plasmon mode is carried out via the excitation of the QD.

In theoretical studies of a metamaterial with compensated losses, it is assumed that electromagnetic wave propagation can be described by the Maxwell equations with real valued effective electric permittivity and magnetic permeability. In its turn, the use of effective permittivities and refraction coefficients assumes that the dipole moment of a NP oscillates with the frequency of the external field and the amplitude of this oscillation is determined by the external field. In the absence of the external field, there should be no oscillations of a dipole moment. The classical linear description treats the gain medium as a system with a negative imaginary part of the permittivity [18-21].

However, when spasers are used as gain inclusions, their dipole moments are excited not only by the external field but also by the radiation produced by QDs, which actually tends to compensate for loss. Unlike the classical linear description [18-21], the semi-classical analysis shows that the spaser in the presence of pumping is a self-oscillating system. The dipole moment of the spaser's NP oscillates autonomously even in the absence of the external field. The autonomic frequency of this self-oscillation is determined by the plasmon frequency, transition frequency of the gain inclusion and characteristic times of relaxation in NP and excitation of gain inclusions [15]. Therefore, in order to develop a correct description of metamaterials with spasers, it is necessary to study the interaction of the spaser with the external field in details.

In this paper, we study the operation of a spaser driven by an external optical wave. We demonstrate that the pumping drastically changes the spaser's behavior in the optical field in comparison with the behavior of a passive QD-NP pair. In particular, when the frequency of the external field is detuned from the autonomous frequency of the spaser, even infinitesimally weak field drives the spaser to stochastic oscillations. The spaser can be synchronized to the external field only when the field amplitude exceeds a threshold value.

The simplest model of the spaser consists of a two-level QD of size $r_{TLS}$ which is positioned at a distance $r$ from a metallic NP of size $r_{NP}$ [11]. The whole system is immersed into a solid dielectric or semiconductor matrix with dielectric permittivity $\varepsilon_M$. The dipole moment of a typical QD of size $r_{TLS} \approx 10$ nm is $\mu_{TLS}^{class} \sim 20$ Debye [22]. For a NP with Ag-core and SiO$_2$-shell with radius $r_{NP} \approx r \approx 10$ nm, the classical dipole moment near the frequency of the plasmon resonance, $\omega$, can be estimated to be $\mu_{NP}^{class} \sim 200$ Debye. The energy of the NP-QD quasistatic dipole-dipole interaction is estimated to be $V = \hbar\Omega_R \sim \mu_{NP}^{class}\mu_{TLS}^{class}/r^3$, which gives $\Omega_R \approx 5\cdot 10^{12} c^{-1} \sim 10^{-2}\omega$. Thus, the Rabi frequency (the coupling constant) is two orders of magnitude smaller than the generation frequency. This allows for the use of the slowly varying envelope approximation in the investigation of the dynamics of the spaser.

The transition processes of the NP and two-level QD interacting in a spaser can be described by the model Hamiltonian

$$\hat{H} = \hat{H}_{SP} + \hat{H}_{TLS} + \hat{V} + \hat{\Gamma}, \tag{1}$$

where

$$\hat{H}_{SP} = \hbar\omega_{SP}\hat{\tilde{a}}^\dagger(t)\hat{\tilde{a}}(t), \tag{2}$$

$$\hat{H}_{TLS} = \hbar\omega_{TLS}\hat{\tilde{\sigma}}^\dagger(t)\hat{\tilde{\sigma}}(t) \tag{3}$$

describe the non-interacting NP and QD, respectively [11, 23, 24], the operator $\hat{V} = -\hat{\boldsymbol{\mu}}_{TLS}\cdot\hat{\mathbf{E}}_{NP}$ determines their interaction. To take into account the interaction of the spaser, which is an open system, with surroundings (pumping, phonons in metal, etc.) following Ref. [24] we introduce the non-Hermitian operator $\hat{\Gamma}$ that describes relaxation and pumping processes. Here $\hat{\tilde{a}}(t)$ is the annihilation operator of the dipole SP, $\hat{\boldsymbol{\mu}}_{TLS} = \boldsymbol{\mu}_{TLS}\left(\hat{\tilde{\sigma}}(t) + \hat{\tilde{\sigma}}^\dagger(t)\right)$ is the operator for the dipole moment of the QD, $\hat{\tilde{\sigma}} = |g\rangle\langle e|$ is the transition operator between ground $|g\rangle$ and excited $|e\rangle$ states of the QD, $\boldsymbol{\mu}_{TLS} = \langle e|e\mathbf{r}|g\rangle$ is the QD dipole moment matrix element. Quantization of the plasmon field without taking into account losses is carried out in the standard way [25]. Thus we obtain [11, 15]

$$\hat{\mathbf{E}}_{NP} = \sum_q \sqrt{\frac{\hbar\omega_q}{2W_q}}\left(\hat{a}_q\mathbf{E}_q(\mathbf{r})\exp(-i\omega_q t) + \hat{a}_q^+\mathbf{E}_q^*(\mathbf{r})\exp(i\omega_q t)\right), \tag{4}$$

where

$$W_q = \frac{1}{4\pi} \int \frac{\partial(\text{Re}\,\varepsilon\omega^2)}{\partial \omega^2}\bigg|_{\omega_q} E_q E_q^* dV = \frac{1}{8\pi} \int \left( \text{Re}\,\varepsilon + \omega \frac{\partial \text{Re}\,\varepsilon}{\partial \omega} \right)\bigg|_{\omega_{SP}} E_1 E_1^* dV \qquad (5)$$

is the normalization factor and $\mathbf{E}_q(\mathbf{r}) = -\nabla \phi_q$ is the $q$-th eigenmode determined by the geometry of the problem [11]. In the dipole approximation, in Eq. (4) we retain only one dipole mode ($q = 1$) with frequency $\omega_{SP} = \omega_1$. For a spherical NP this dipole mode is uniform inside the particle and has the form of the field of the dipole with a unitary dipole moment outside of the NP: $\mathbf{E}_1 = -\frac{\mathbf{e}}{r^3} + 3\frac{(\mathbf{e}\cdot\mathbf{r})\mathbf{r}}{r^5}$. Since for the field of the dipole $\frac{1}{8\pi} \int \text{Re}\,\varepsilon\big|_{\omega_{SP}} E_1 E_1^* dV = 0$ [26] and $\partial \varepsilon / \partial \omega = 0$ outside of the particle, we obtain $W_1 = \frac{1}{8\pi} \int_{\substack{volume \\ of\ NP}} \omega \frac{\partial \text{Re}\,\varepsilon}{\partial \omega}\bigg|_{\omega_{SP}} E_1 E_1^* dV = \frac{r_{NP}^3}{6} \omega \frac{\partial \text{Re}\,\varepsilon}{\partial \omega}\bigg|_{\omega_{SP}}$. This gives $\hat{\mathbf{E}}_{NP} = \sqrt{3\hbar/(r_{NP}\partial \varepsilon/\partial \omega)}\,\mathbf{E}_1(\mathbf{r})(\hat{\tilde{a}} + \hat{\tilde{a}}^+)$. Thus, the operator for the dipole moments of the NP is $\hat{\boldsymbol{\mu}}_{NP} = \boldsymbol{\mu}_{NP}(\hat{\tilde{a}} + \hat{\tilde{a}}^\dagger)$, where $\boldsymbol{\mu}_{NP} = \sqrt{3\hbar r_{NP}^5/(\partial \text{Re}\,\varepsilon_M/\partial \omega)}$.

Assuming that the frequencies of the QD transition and the frequency of the dipole SP are close, $\omega_{SP} \approx \omega_{TLS}$, we will look for the time dependencies of $\hat{\tilde{a}}(t)$ and $\hat{\tilde{\sigma}}(t)$ in the form $\hat{\tilde{a}}(t) = \hat{a}(t)\exp(-i\omega_a t)$ and $\hat{\tilde{\sigma}}(t) = \hat{\sigma}(t)\exp(-i\omega_a t)$, where $\hat{a}(t)$ and $\hat{\sigma}(t)$ are the slowly changing amplitudes and $\omega_a$ is the autonomous frequency of the spaser which we seek. In the rotating wave approximation [24] we can neglect rapidly oscillating terms, $\propto \exp(\pm 2i\omega_a t)$, and obtain the interaction operator $\hat{V}$ in the form of the Jaynes–Cummings Hamiltonian [27]

$$\hat{V} = \hbar \Omega_R (\hat{a}^\dagger \hat{\sigma} + \hat{\sigma}^\dagger \hat{a}). \qquad (6)$$

where Rabi frequency is $\Omega_R = \sqrt{3/(\hbar r_{NP} \partial \text{Re}\,\varepsilon/\partial \omega)}\,\mathbf{E}_1(\mathbf{r}) \cdot \boldsymbol{\mu}_{\mathbf{eg}}$.

The commutation relations for operators $\hat{a}(t)$ and $\hat{\sigma}(t)$ are standard: $[\hat{a}, \hat{a}^\dagger] = \hat{1}$ and $[\hat{\sigma}^\dagger, \hat{\sigma}] = \hat{D}$, where the operator $\hat{D}$ describes the inversion of the occupancies of the ground and excited states of the QD, $\hat{D}(t) = \hat{n}_e(t) - \hat{n}_g(t)$, $\hat{n}_e = |e\rangle\langle e|$ and $\hat{n}_g = |g\rangle\langle g|$ are the occupancy operators for the ground and excited states, $\hat{n}_g + \hat{n}_e = \hat{1}$. Using the Hamiltonian, Eqs. (1)-(3), and the commutation relations for operators $\hat{a}(t)$ and $\hat{\sigma}(t)$ we obtain the Heisenberg equations of motion for the operators $\hat{a}(t)$, $\hat{\sigma}(t)$, and $\hat{D}(t)$ [28, 29]:

$$\dot{\hat{D}} = 2i\Omega_R(\hat{a}^\dagger\hat{\sigma} - \hat{\sigma}^\dagger\hat{a}) - \frac{\hat{D} - \hat{D}_0}{\tau_D}, \tag{7}$$

$$\dot{\hat{\sigma}} = (i\delta_{TLS} - \frac{1}{\tau_\sigma})\hat{\sigma} + i\Omega_R\hat{a}\hat{D}, \tag{8}$$

$$\dot{\hat{a}} = (i\delta_{SP} - \frac{1}{\tau_a})\hat{a} - i\Omega_R\hat{\sigma}, \tag{9}$$

where $\delta_{TLS} = \omega_a - \omega_{TLS}$ and $\delta_{SP} = \omega_a - \omega_{SP}$ are the frequency differences. The Markovian interaction with the reservoir is described by the Liouvillian (non-Hermitian operator) term $\hat{\Gamma}$ determining the decay rates with relaxation times $\propto \tau_D^{-1}, \tau_\sigma^{-1}$, and $\tau_a^{-1}$, which account for the relaxation processes for the SP annihilation operator, the QD polarization and the occupancy operators respectively [27]. The operator for the occupancy inversion $\hat{D}_0$ describes the pumping [24, 27]. We now neglect quantum fluctuations and correlations and consider $\hat{a}(t)$, $\hat{\sigma}(t)$ and $\hat{D}(t)$ as complex quantities (c-numbers), so that we can use complex conjugation instead of Hermitian conjugation [11, 29-31]. The difference between the occupancies of the upper and lower levels $D(t)$ must be a real valued quantity because the respective operator is Hermitian. The quantities $\sigma(t)$ and $a(t)$ are the complex amplitudes of the dipole oscillations of the QD and SP, respectively.

The system of equations (7)-(9) has two stationary solutions. The trivial unstable solution corresponds to the absence of SPs, while the stable one corresponds to laser generation:

$$a = \frac{e^{i\varphi}}{2}\sqrt{\frac{(D_0 - D_{th})\tau_a}{\tau_D}}, \quad \sigma = \frac{e^{i\psi}}{2}\sqrt{\frac{(D_0 - D_{th})(\delta_{SP}^2 + 1/\tau_a^2)\tau_a}{\Omega_R^2\tau_D}},$$

$$D = D_{th}, \quad \omega = \frac{\omega_{SP}\tau_a + \omega_{TLS}\tau_\sigma}{\tau_a + \tau_\sigma}, \tag{10}$$

where $\exp(i(\psi-\varphi))=(\delta_{SP}+i/\tau_a)/\sqrt{\delta_{SP}^2+1/\tau_a^2}$. The unstable solution appears only when the pumping level reaches the threshold value $D_{th}=(1+\delta_{SP}^2\tau_a^2)/(\Omega_R^2\tau_a\tau_\sigma)$ (see Fig. 1). The stationary values of $a$, $\sigma$, and $D$ shown in Fig. 1.

Fig. 1 (Color online) Stationary amplitudes $a$ and $\sigma$ are shown by dash-double dotted and dash-dotted lines, respectively. The stable solution for $D$ is shown by the solid line. The unstable solution appearing for $D>D_{th}$ is shown by the dashed line. For $D_0=D_0'$ (also shown in Fig. 2) the stable and unstable solutions of $D$ are marked by red dots.

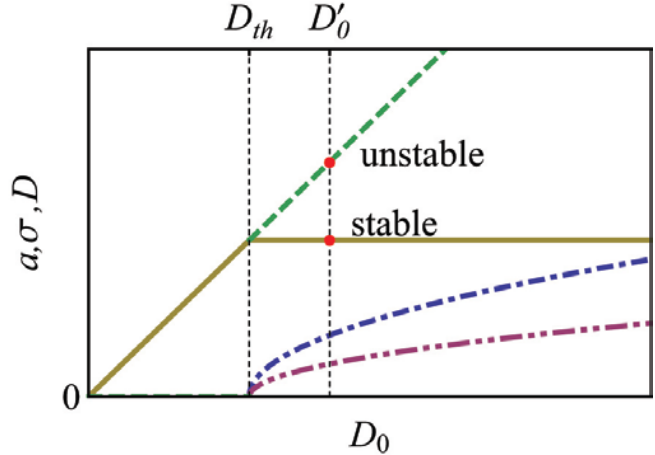

Let us consider the dynamics of the NP and QD in the field of the external optical wave, $E(t)=E\cos\omega_f t$. Assuming that the external field is classical and taking into account the dipole interaction only, we can write the Hamiltonian in the form

$$\hat{H}_{ef}=\hat{H}+\hbar\Omega_1(\hat{a}^\dagger+\hat{a})(\exp(i\omega_f t)+\exp(-i\omega_f t))+\hbar\Omega_2(\hat{\sigma}^\dagger+\hat{\sigma})(\exp(i\omega_f t)+\exp(-i\omega_f t)),$$

where $\hat{H}$ is defined by Eq. (2), $\Omega_1=-\mu_{NP}E/\hbar$, and $\Omega_2=-\mu_{TLS}E/\hbar$. As before, we use the Heisenberg equations for operators $\hat{a}$, $\hat{\sigma}$, and $\hat{D}$ to obtain equations of motion for "slowly varying" amplitudes,

$$\dot{\hat{D}}=2i\Omega_R(\hat{a}^\dagger\hat{\sigma}-\hat{\sigma}^\dagger\hat{a})+2i\Omega_2(\hat{\sigma}-\hat{\sigma}^\dagger)-\frac{\hat{D}-\hat{D}_0}{\tau_D}, \qquad (11)$$

$$\dot{\hat{\sigma}}=(i\Delta_{TLS}-\frac{1}{\tau_\sigma})\hat{\sigma}+i\Omega_R\hat{a}\hat{D}+i\Omega_2\hat{D}, \qquad (12)$$

$$\dot{\hat{a}} = (i\Delta_{SP} - \frac{1}{\tau_a})\hat{a} - i\Omega_R \hat{\sigma} - i\Omega_1. \tag{13}$$

Below we assume for simplicity that $\omega_{TLS} = \omega_{SP}$. Hence $\Delta_{SP} = \omega_f - \omega_{SP} = \Delta_{TLS} = \omega_f - \omega_{TLS} = \Delta$. Note that without the QD, the polarization of the NP corresponding to the stationary solution of the system of equations (11)-(14), $\dot{\hat{a}} = \dot{\hat{\sigma}} = 0$, has the form

$$\alpha r_{NP}^{-3} = \frac{3\hbar r_{NP}^5}{i\Delta_{SP} - 1/\tau_a} \left(\frac{\partial \varepsilon_{NP}}{\partial \omega}\right)^{-1},$$

where $\varepsilon_{NP}$ is the permittivity of the metal NP. In the slowly varying amplitude approximation, i.e. for a small detuning $\Delta_{SP} \ll 1$, this expression coincides with the classical one $\alpha^{class} r_{NP}^{-3} = \frac{\varepsilon_{NP}(\omega) - \varepsilon_M}{\varepsilon_{NP}(\omega) + 2\varepsilon_M}$ (see also Ref. [30]) .

Note that quantities defined by Eqs. (11)-(13) differ from those in Eqs. (7)-(9). Without the external field the frequency $\omega_a$ in the substitutions $\hat{\tilde{a}}(t) = \hat{a}(t)\exp(-i\omega_a t)$ and $\hat{\tilde{\sigma}}(t) = \hat{\sigma}(t)\exp(-i\omega_a t)$ is found from the solution of stationary equations (7)-(9). When the external field is present, we use the representations $\hat{\tilde{a}}(t) = \hat{a}(t)\exp(-i\omega_f t)$ and $\hat{\tilde{\sigma}}(t) \equiv \hat{\sigma}(t)\exp(-i\omega_f t)$. We look for a stationary solution of the system of equations (11)-(13). The stationary values of $D$ satisfy a cubic equation and are shown in Fig. 2 as a function of the amplitude of the external field $E$. For $\Delta = 0$ (Fig. 2a) in the absence of the external field $E = 0$ and $D_0 \geq D_{th}$ two points, stable and unstable, correspond to the points marked by dots in Fig. 1.

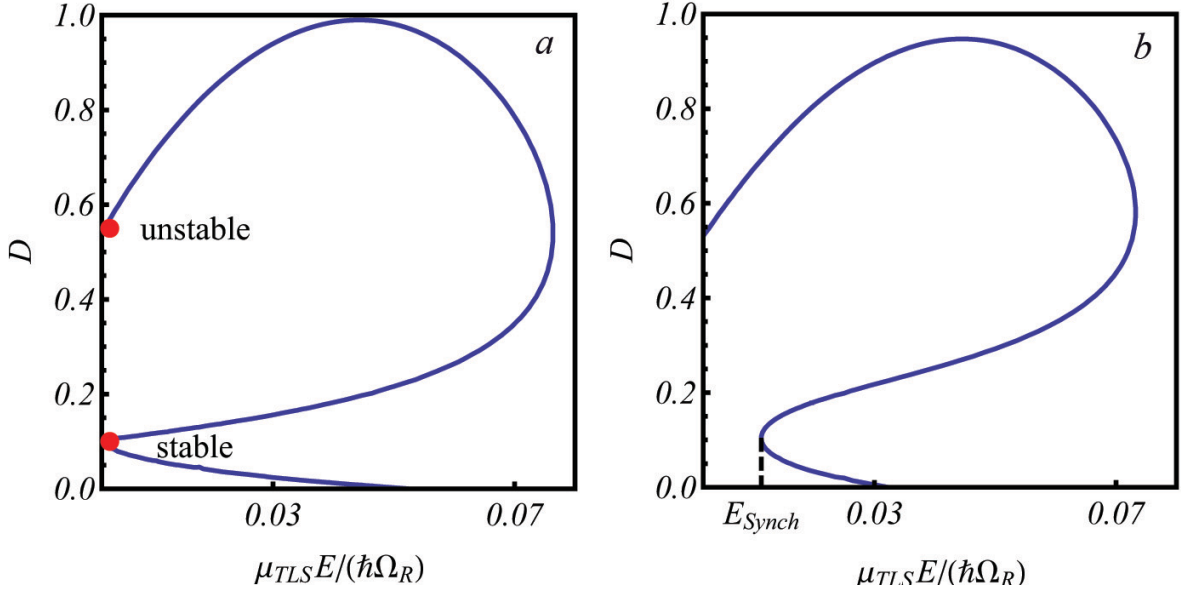

Fig. 2. (Color online) The stationary values of $D$ as a function of the amplitude of the external field for (*a*) zero ($\Delta = 0$) and (*b*) non-zero ($\Delta = 10^{11} s^{-1}$) detuning. For both graphs $\tau_a = 10^{-14} s$, $\tau_\sigma = 10^{-11} s$, $\tau_D = 0.5 \cdot 10^{-14} s$, $\Omega_R = 10^{13} s^{-1}$, $D_0 = D_0' = 0.55$.

Only the lower branch of the synchronization region corresponds to a stable solution. The stationary solutions at zero external field and with no detuning ($\delta_{TLS} = \delta_{SP} = \Delta = 0$) are shown in Fig. 1. When detuning is present (we still assume that $\delta_{SP} = \delta_{TLS} = 0$ but $\Delta \neq 0$), the stable stationary solution exists for $E > E_{Synch}(\Delta)$ only (Fig. 2b). This result is confirmed by our numerical simulation shown in Fig. 3. Such behavior of a self-oscillating system is referred to as synchronization by an external periodic influence [32]. The region in which the synchronization takes place is known as the Arnold tongue.

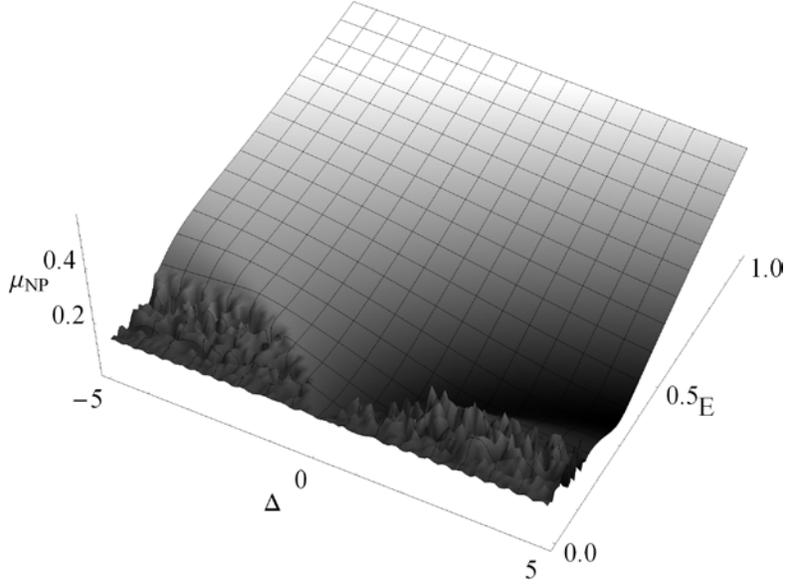

Fig. 3. The dependence of the plasmon dipole moment on the amplitude of the external field and the frequency detuning $\Delta$. The speckle structure at low values of $E$ corresponds to the chaotic behavior of the dipole moment.

To obtain an analytical estimate of the Arnold tongue boundary, we find the solution of Eqs. (11)–(13) in the first approximation with respect to the field. For the case $\delta_{SP} = 0$, using the substitution $a = |a|\exp(i\varphi)$ and $\sigma = |\sigma|\exp(i\psi)$ in Eq. (13) and equating its imaginary parts, we obtain

$$\dot{\varphi} = \Delta - \Omega_R \frac{|\sigma|}{|a|}\cos(\psi - \varphi) - \frac{\Omega}{|a|}\cos\varphi.$$

For $|a|$, $|\sigma|$, and $\cos(\psi - \varphi)$ we can use their values obtained from Eq. (10) in the zeroth order with respect to the field approximation. As the result, we have

$$\dot{\varphi} = \Delta - \xi\cos\varphi, \tag{14}$$

where $\xi = \Omega_1 / |a|$ (in this approximation $\cos(\psi - \varphi) = 0$). If one introduces a function $U(\varphi) = \Delta\left(-\varphi + \frac{\xi}{\Delta}\sin\varphi\right)$, Eq. (14) takes the form of Adler's equation [32]:

$$\dot{\varphi} = -\frac{\partial U(\varphi)}{\partial \varphi}.$$

The phase dynamics can be viewed as the motion of a particle sliding along the potential profile $U(\varphi)$ (Fig. 4) in a viscous liquid. For a small field and/or large detuning, $|\xi/\Delta|<1$, the phase difference of the system and the external field increases monotonously. For $|\xi/\Delta|>1$, the "particle" should be trapped in one of the minima of the potential function. This corresponds to the regime of synchronization: the phase of the system oscillations does not depend upon time. The Arnold tongue is the set of parameters $\xi$ and $\Delta$ for which synchronization occurs. In the approximation considered, the width of this region is proportional to the field amplitude. Within the Arnold tongue, for small fields of the external wave (the energy of the interaction of the NP with the external field is smaller than the QD-NP interaction), the amplitude of the auto-oscillations depends weakly upon the amplitude of the external field, which in this case plays the role of a synchronizer.

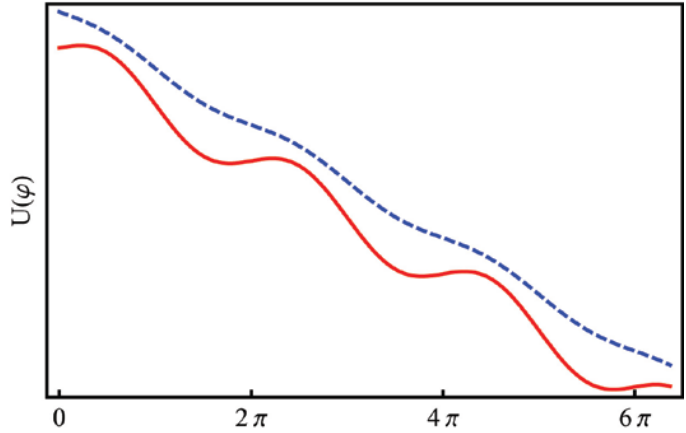

Fig. 4. (Color online) The potential $U(\varphi)$ for $|\xi/\Delta|<1$ (dashed line) and $|\xi/\Delta|>1$ (solid line).

When $\Delta \to 0$, the threshold field value $E_{Synch}$ is determined by the equation $(\mu_{NP} E_{Synch}/\hbar)^2 = (D_0 - D_{th})\Delta^2(\tau_a/4\tau_D)$. For big values of detuning, $E_{Synch}$ becomes independent of $\Delta$ (see Fig.3). Taking into account that for plasmonic NPs $\tau_a \sim 10^{-14} s^{-1}$ and $\tau_\sigma \sim 10^{-11} s^{-1}$ one can estimate the asymptotic value of the Arnold tongue boundary, $E^*_{Synch}$. Our numerical calculations show that $E_{Synch}(\Delta)$ tends to a plateau for $\Delta \sim 2 \cdot 10^{11} s^{-1}$, which gives $E^*_{Synch} \sim 3 \cdot 10^3 V/m$. This agrees with calculations of Refs. [3, 10] in which the wave propagation at the system of spasers was considered. In Refs. [3, 10] the amplitude of the incident wave was by a few orders of magnitude greater than $E^*_{Synch}$. Such a field is comparable with the near field inside the spaser. In this case, spaser synchronizes with the external field for any value of the detuning; it ceases to be an autonomous system and responses linearly to the external field as can be seen in Fig. 3. For such strong fields there is no compensation of losses and only amplification can occur.

Authors are indebted to Yu.E. Lozovik and C. Z. Ning for useful discussions. This work was supported by RFBR Grants Nos. 10-02-91750 and 11-02-92475 and PSC-CUNY grant.

**Figure Captions**

Fig. 1 (Color online) Stationary amplitudes $a$ and $\sigma$ are shown by dash-double dotted and dash-dotted lines, respectively. The stable solution for $D$ is shown by the solid line. The unstable solution appearing for $D > D_{th}$ is shown by the dashed line. For $D_0 = D_0'$ (also shown in Fig. 2) the stable and unstable solutions of $D$ are marked by red dots.

Fig. 2. (Color online) The stationary values of $D$ as a function of the amplitude of the external field for (a) zero ($\Delta = 0$) and (b) non-zero ($\Delta = 10^{11} s^{-1}$) detuning. For both graphs $\tau_a = 10^{-14} s$, $\tau_\sigma = 10^{-11} s$, $\tau_D = 0.5 \cdot 10^{-14} s$, $\Omega_R = 10^{13} s^{-1}$, $D_0 = D_0' = 0.55$.

Fig. 3. The dependence of the plasmon dipole moment on the amplitude of the external field and the frequency detuning $\Delta$. The speckle structure at low values of $E$ corresponds to the chaotic behavior of the dipole moment.

Fig. 4. (Color online) The potential $U(\varphi)$ for $|\xi/\Delta| < 1$ (dashed line) and $|\xi/\Delta| > 1$ (solid line).